# Local Electron Field Emission Study of Two-dimensional Carbon

Ying Wang[1], Yumeng Yang[1], Zizheng Zhao[1], Chi Zhang[1,2], and Yihong Wu[1,*]

[1]Information Storage Materials Laboratory, Department of Electrical and Computer Engineering, National University of Singapore, 4 Engineering Drive 3, Singapore 117583

[2]NUS Graduate School for Integrative Sciences and Engineering, Centre for Life Sciences (CeLS), #05-01, 28 Medical Drive, Singapore 117456

A systematic field-emission study has been carried out on different types of two-dimensional carbons in ultrahigh vacuum with the cathode-anode distance ranging from near-contact to about 124 nm. An analytical model has been developed to explain the increase of field enhancement factor with the cathode-anode distance. Good agreement has been achieved between the calculation results and experimental data, including those reported in literature. The work provides useful insights into the characteristics of field emission from two-dimensional carbon.

* Author to whom correspondence should be addressed: elewuyh@nus.edu.sg

Two-dimensional (2D) carbon has attracted attention as promising electron field emitters due to its large field enhancement factor (β) stemming from its unique shape and dimensions.[1-3] Despite intensive investigations both theoretically and experimentally, however, the exact range of values for β and its quantitative dependence on the dimensions of 2D carbon and anode-cathode distance (d) are still debatable. The experimental values for β extracted from fittings to Fowler-Nordheim (F-N) plot[4] range from $10^3$ to $3 \times 10^4$, based on experiments conducted on a variety of 2D carbons with d ranging from 20 to 1000 μm.[1,2,5-35] Several theoretical studies have revealed that β is largely determined by the height to thickness ratio of 2D carbon.[36-38] Although these models are in qualitative agreement with experimental observations, the calculated values of β are at least one order of magnitude smaller than the experimental values and, in addition, a satisfactory explanation of the d-dependence of β has yet to be obtained.[36-38] Considering the importance of β in understanding the field emission mechanism of 2D carbon, it is of crucial importance that additional data can be obtained from experiments conducted in an ultra-clean environment and using an experimental setup that allows for variation of cathode-anode distance from near contact to the sub-micron regime with nanometer accuracy. Furthermore, it will be desirable to develop an analytical model that is able to account for the experimental results obtained so far on β both in the value and its dependence on the sample dimensions and anode-cathode distance.

In view of the above, in this letter, we first describe our systematic study on the field emission of different types of 2D carbons in an ultrahigh vacuum (UHV) nanoprobe system which allows one to control the anode-cathode distance at nanometer accuracy in a range from ~1 nm to 124.2 nm. The experimental results will be compared with the reported data in literature. In order to explain the experimental observations, we further introduce an analytical model based on basic electrostatics. Good agreement between the theoretical model and experimental results, including those reported in literature and our own results, has been obtained for both the value and d-dependence of β.

Three different types of 2D samples, namely carbon nanowalls on $SiO_2$ (CNW/$SiO_2$), CNW on Cu (CNW/Cu) and exposed edges of chemical vapor deposited (CVD) single-layer graphene on Cu (SLG/Cu), were employed for the current study. The CNW was chosen because it stands rigidly and vertically on a substrate and thus best suits the present study.[39] The SLG/Cu sample is used for comparison with other reported experimental results. Details about the CNW growth and structure can be found elsewhere.[1,40] The SLG/Cu sample was prepared by first "cutting-and-tearing" a SLG/Cu strip with a STM tip plier, followed by "dip-etching" the freshly exposed edge in 1M $Fe(NO_3)_3$ solution for 14 min and then "dip-rinsing" in DI water for 10 min. The sample was then fastened between two pieces of Si wafers (to make the graphene sheet stand vertically on the sample holder) and loaded into vacuum chamber immediately after the preparation [see inset of Fig. 1(c) for an illustration].

All investigations were subsequently carried out in an Omicron UHV system with a base pressure better than $2.2 \times 10^{-10}$ mbar. Equipped inside the UHV system are a scanning electron microscope (SEM) and four independently controllable nanoprobes with auto-approaching capability, which allow for position-specific measurements down to nanoscale. Figures 1(b) and (d) are the schematic diagrams of the measurement setup.

Prior to field emission (FE) measurements, the average step size of the W anode has been carefully calibrated to be ~1.38 nm by using patterned gold pads with pre-determined height. The stability of the experiment setup has been confirmed by monitoring the emission current by applying a constant voltage bias for a duration which is much longer than the time needed for each round of FE measurement. The FE measurements always started with preparation of the W anode into desired size (0.6 – 2 µm) and shape through local electrical melting inside the chamber by applying a current of appropriate magnitude between the tip apex of the anode probe and the body of another W probe that was firmly pressed onto 2D carbon to form a close-loop for FE measurements. This process is critical to obtain a sub-micron to micron sized anode with a smooth surface which is in turn crucial for FE measurements with good reproducibility. The W anode was then carefully approached to a 2D carbon single flake through monitoring the differential contact resistance using a lock-in amplifier setup. After the electrical contact was achieved, the tip was subsequently lifted with pre-calibrated steps (~1.38 nm/step) for FE measurements at different distances (d) ranging from 1.38 nm to 124.2 nm [Fig. 1(a) and Fig. 1(c)]. Multiple runs of FE measurements were performed at each d in order to improve the reliability of the measurement results. All the measurements were carried out in the UHV chamber at room temperature. Before proceeding to discuss the experimental results, it should be emphasized that the FE properties revealed in this work originate from local FE of 2D carbon, which has been confirmed by performing FE measurements with W anode of different sizes (0.6 – 2 µm).

Figure 2(a) shows the emission current – electric field (I-E) plot obtained from the SLG/Cu sample with a 2 µm-in-diameter W tip at different d; the corresponding F-N plot is shown in Fig. 2(b) and (c). Figure 2(d) and (e) is the F-N plot from the CNW/Cu and CNW/SiO$_2$ sample, respectively. Multiple (3-8) sets of curves are displayed in the figures for each d; close overlap of the curves at each d indicates good reproducibility of the experimental data. It has been found that most of the F-N curves of all three types of samples exhibit good linearity, in agreement with the F-N model. Although it was also noticed that the F − N curves at small and large distances show reproducible gentle superlinear and sublinear characteristics, respectively, the origin will be discussed elsewhere. Figure 3(a) plots the averaged β (symbols) extracted from the slope of the F-N curves in Fig. 2(b) - (e) using the relation:[41]

$$\ln(\frac{I}{E^2}) = \ln(\frac{SA\beta^2}{\Phi}) - \frac{B\Phi^{3/2}}{\beta}\frac{1}{E},$$
(1)

where A = 1.54×10$^{-6}$ AV$^{-2}$eV, B = 6.83×10$^3$ eV$^{-3/2}$Vμm$^{-1}$, Φ = 5 eV and S is the emission area. Each data point is obtained by averaging all the calculated β from multiple rounds of measurement at the same distance. The most important observation is that β increases rapidly with increasing d for all three types of 2D samples. This β-d relation in turn results in the strong dependence of the turn-on field (defined here as field for obtaining an emission current of 1 nA) on d, which can be seen in Fig. 3(b). The data at large d (unfilled symbols) are gathered from literature,[1,2,5-20,22-35,42-45] whereas those at small d (filled symbols) are the results of this study. As discussed in the introduction, theoretic studies reported so far, both numerical modeling and analytical calculations employing conformal mapping, have failed to achieve a good agreement with experimental results in both the value of β and its dependence on d. Here, we propose a simple analytical model based on basic electrostatics to assist in understanding the behavior of β.

The inset of Fig. 3(c) shows the schematic of the model for calculating β. The black rectangle and length extending infinitely in y-axis represents the 2D carbon emitter. To facilitate analytical calculation, the tungsten (W) anode is modeled as a flat plate placed at an infinite d above the upper edge of 2D carbon and extending infinitely in the x-y plane. The W plate is grounded and a voltage bias V is applied to the 2D carbon. As the height to thickness ratio of 2D carbon is very large and in order to simplify the calculation, we assume that the electric field has a vertical component on the top surface of 2D carbon and is zero elsewhere. Although this assumption may lead to an overestimation of β, it is adequate for describing the spatial variation of electric field in the vertical direction. Based on this assumption, the electric field at z = 0 is given by

$$E_z(x,0) = \begin{cases} E_l & if -t/2 \leq x \leq t/2 \\ 0 & otherwise \end{cases}. \qquad (2)$$

Here, $E_l$ is the local field and t is the thickness of 2D carbon. Based on these assumptions, the electric field in the free space between the 2D carbon and W anode can be obtained by solving the two-dimensional Laplace equation satisfying the above-mentioned boundary conditions, in analogy to a magnetic writer head:[46]

$$\begin{aligned} E_x(x,z) &= -\frac{E_l}{2\pi} \ln \frac{(t/2+x)^2 + z^2}{(t/2-x)^2 + z^2} \\ E_z(x,z) &= \frac{E_l}{\pi} [\arctan(\frac{t/2+x}{z}) + \arctan(\frac{t/2-x}{z})] \quad (z \geq 0) \end{aligned}. \qquad (3)$$

β can be calculated analytically as follows:

$$\beta = \frac{E_l}{(1/d)\int_0^d E_z(0,z)dz} = \frac{2\pi}{(t/d)\ln[1+4(d/t)^2]+4\arctan[(1/2)(t/d)]}. \quad (4)$$

We first look at the extreme case when $(d/t) \rightarrow \infty$, which gives $\beta = 1.4\pi(d/t)$. This simple yet intriguing result can be readily understood as follows. $2\pi/t$ is the characteristic spatial frequency of the electrical field distribution along x-axis at z = 0. This spatial frequency determines the rate of exponential decay of electrical potential in the z-direction from the 2D carbon surface. Therefore, the local electric field is approximately given by $2\pi V/t$ with V being the electric potential of 2D carbon surface. As the global field is V/d, one can readily obtain an enhancement factor of $2\pi d/t$, which is close to $1.4\pi(d/t)$.

Figure 3(c) shows the calculated dependence of β on normalized distance (d/t) ranging from 1 to $1\times10^6$ at x = 0 (dotted-line) in double-logarithmic scale. The calculation result shows that β increases with increasing d/t for d >> t, and has a value of about 24000 at d/t = $1\times10^5$. Here we first look at how the theoretical values are compared with the experimental data at large d [Fig. 3(b)]. As the local electric field is presumably the same as the global field as $d \rightarrow 0$, the ratio between the average turn-on field at large d (dotted-line) and that when $d \rightarrow 0$ should be the experimental enhancement factor, which turned out to be around $1.2\times10^4$ at d = $1\times10^5$ nm. Since the typical thickness of 2D carbon is within a few nm, this is in good agreement with the calculated values $5.42\times10^3$ $- 2.38\times10^4$ for t = 1 − 5 nm at d = $1\times10^5$ nm from the model. The variation in the reported turn-on field at large d in Fig. 3(b) can be understood as being caused by the variation of carbon thickness from sample to sample in different sets of experiments reported in literature.

Although Eqs. (3) and (4) were obtained by assuming d → ∞, the results should be equally valid at a much smaller d as exp(-2πd/t) << 1 when d > 2t. We now compare the calculation result with experimental result at small d. High-resolution transmission electron microscopy (HRTEM) observations revealed that the CNW are vertically aligned graphene sheets with thickness ranging from one to several nanometers.[1,47] However, it is practically very challenging to examine the exact thickness of the particular 2D carbon flake under investigation since it is difficult to distinguish it from the rest after taking out the sample (1 cm$^2$) from the measurement chamber. We have thus calculated β with t = 2 nm which is close to the median value of 2D carbon thickness for a qualitative comparison [solid curve in Fig. 3(a)]. It can be seen that the model is able to reproduce the general trend of the d-dependence of β. However, it predicts a larger β by a factor of 4 to 10 in the range of 1.4 nm ≤ d ≤ 124 nm. This overestimation in the value of β is resulted from assuming that only the top surface of 2D carbon has a vertical field. Despite the adequateness of this assumption for the case of large d/t, however, the electric field becomes less concentrated on the top surface of the 2D carbon at d comparable to t. The change in the local distribution of electric field in turn results in a smaller β than the calculation result. To justify our argument, we further performed finite element analysis to obtain the field distribution around 2D carbon at d/t = 1, which took into account of the electric field originates from the sidewall of 2D carbon. In the simulation, the same geometry model as our analytical calculation has been used [inset of Fig. 3(c)]. The applied voltage was set to 30 V which is the typical value used in our experiments. The height-to-thickness ratio and relative permittivity of 2D carbon was assumed to be 5 and 10, respectively. The inset of Fig. 3(a) shows simulated z-component of the total field strength normalized by the global field (i.e. V/d) at the gap region. Clearly, the local field at the top surface of the 2D carbon is smaller than the global field, suggesting an enhancement factor less than unity.

In conclusion, we have shown that the enhancement factor of 2D carbon emitter is determined by the ratio between the sample-anode distance and thickness of 2D carbon through analytical calculation based on a simple electrostatic model. Good agreement has been achieved between the calculation results and experimental data, including both the data obtained in this study and those reported in literature. The enhancement factor at small cathode-anode distance was found to be smaller than unity due to the change of the local distribution of electric field at the 2D carbon emitter surface. This work provides some useful insights into the characteristics of FE from 2D carbon.

We wish to thank B. L. Wu and Z. X. Chen for some helpful discussions. This work is supported by the National Research Foundation of Singapore (Grants No. NRF-G-CRP 2007-05 and R-143-000-360-281).

References

[1]  Y. H. Wu, B. J. Yang, B. Y. Zong, H. Sun, Z. X. Shen, and Y. P. Feng, J. Mater. Chem. **14**, 469 (2004).
[2]  S. G. Wang, J. J. Wang, P. Miraldo, M. Y. Zhu, R. Outlaw, K. Hou, X. Zhao, B. C. Holloway, D. Manos, T. Tyler, O. Shenderova, M. Ray, J. Dalton, and G. McGuire, Appl. Phys. Lett. **89**, 183103 (2006).
[3]  M. Hiramatsu and M. Hori, *Carbon Nanowalls* (Springer-Wien, New York, 2010), p. 31.
[4]  R. H. Fowler and L. Nordheim, Proc. R. Soc. Lond. A **119**, 173 (1928).
[5]  A. N. Obraztsov, I. Y. Pavlovsky, and A. P. Volkov, J. Vac. Sci. Technol. B **17**, 674 (1999).

6  A. N. Obraztsov, I. Y. Pavlovsky, A. P. Volkov, A. S. Petrov, V. I. Petrov, E. V. Rakova, and V. V. Roddatis, Diam. Relat. Mater. **8**, 814 (1999).

7  R. Kurt, J. M. Bonard, and A. Karimi, Carbon **39**, 1723 (2001).

8  A. N. Obraztsov, A. P. Volkov, K. S. Nagovitsyn, K. Nishimura, K. Morisawa, Y. Nakano, and A. Hiraki, J. Phys. D. Appl. Phys. **35**, 357 (2002).

9  J. J. Wang, M. Y. Zhu, X. Zhao, R. A. Outlaw, D. M. Manos, B. C. Holloway, C. Park, T. Anderson, and V. P. Mammana, J. Vac. Sci. Technol. B **22**, 1269 (2004).

10  J. Y. Wang, T. Teraji, and T. Ito, Diam. Relat. Mater. **14**, 2074 (2005).

11  T. Itoh, S. Shimabukuro, S. Kawamura, and S. Nonomura, Thin Solid Films **501**, 314 (2006).

12  J. Y. Wang and T. Ito, Diam. Relat. Mater. **16**, 589 (2007).

13  J. Y. Wang and T. Ito, Diam. Relat. Mater. **16**, 364 (2007).

14  M. Y. Chen, C. M. Yeh, J. S. Syu, J. Hwang, and C. S. Kou, Nanotechnology **18**, 185706 (2007).

15  A. T. H. Chuang, J. Robertson, B. O. Boskovic, and K. K. K. Koziol, Appl. Phys. Lett. **90**, 123107 (2007).

16  A. Malesevic, R. Kemps, A. Vanhulsel, M. P. Chowdhury, A. Volodin, and C. Van Haesendonck, J. Appl. Phys. **104**, 084301 (2008).

17  G. R. Gu and T. Ito, Diam. Relat. Mater. **17**, 817 (2008).

18  M. Bagge-Hansen, R. A. Outlaw, P. Miraldo, M. Y. Zhu, K. Hou, N. D. Theodore, X. Zhao, and D. M. Manos, J. Appl. Phys. **103**, 014311 (2008).

19  Z. S. Wu, S. F. Pei, W. C. Ren, D. M. Tang, L. B. Gao, B. L. Liu, F. Li, C. Liu, and H. M. Cheng, Adv. Mater. **21**, 1756 (2009).

20  M. Qian, T. Feng, H. Ding, L. F. Lin, H. B. Li, Y. W. Chen, and Z. Sun, Nanotechnology **20**, 425702 (2009).

21  W. T. Zheng, Y. M. Ho, H. W. Tian, M. Wen, J. L. Qi, and Y. A. Li, J. Phys. Chem. C **113**, 9164 (2009).

22  W. C. Shih, J. M. Jeng, C. T. Huang, and J. T. Lo, Vacuum **84**, 1452 (2010).

23  V. I. Kleshch, T. Susi, A. G. Nasibulin, E. D. Obraztsova, A. N. Obraztsov, and E. I. Kauppinen, Phys. Status Solidi B **247**, 3051 (2010).

24  J. L. Liu, B. Q. Zeng, Z. Wu, J. F. Zhu, and X. C. Liu, Appl. Phys. Lett. **97**, 033109 (2010).

25  U. A. Palnitkar, R. V. Kashid, M. A. More, D. S. Joag, L. S. Panchakarla, and C. N. R. Rao, Appl. Phys. Lett. **97**, 063102 (2010).

26  G. R. Gu and T. Ito, Appl. Surf. Sci. **257**, 2455 (2011).

27  C. K. Huang, Y. X. Ou, Y. Q. Bie, Q. Zhao, and D. P. Yu, Appl. Phys. Lett. **98**, 263104 (2011).

28  M. Y. Zhu, R. A. Outlaw, M. Bagge-Hansen, H. J. Chen, and D. M. Manos, Carbon **49**, 2526 (2011).

29  Q. S. Huang, G. Wang, L. W. Guo, Y. P. Jia, J. J. Lin, K. Li, W. J. Wang, and X. L. Chen, Small **7**, 450 (2011).

30  Q. Zhao, T. Cai, Y. Ou, C. Huang, R. Zhu, Y. Sun, Y. Zhou, Z. Liao, H. Peng, and D. Yu, Adv. Sci. Lett. **5**, 192 (2012).

31  Z. C. Yang, Q. Zhao, Y. X. Ou, W. Wang, H. Li, and D. P. Yu, Appl. Phys. Lett. **101**, 173107 (2012).

32  E. Stratakis, G. Eda, H. Yamaguchi, E. Kymakis, C. Fotakis, and M. Chhowalla, Nanoscale **4**, 3069 (2012).

33  J. L. Liu, B. Q. Zeng, Z. Wu, and H. Sun, ACS Appl. Mater. Interfaces **4**, 1219 (2012).

34  J. H. Deng, R. T. Zheng, Y. Zhao, and G. A. Cheng, Acs Nano **6**, 3727 (2012).

35  L. L. Jiang, T. Z. Yang, F. Liu, J. Dong, Z. H. Yao, C. M. Shen, S. Z. Deng, N. S. Xu, Y. Q. Liu, and H. J. Gao, Adv. Mater. **25**, 250 (2013).

36  R. Miller, Y. Y. Lau, and J. H. Booske, Appl. Phys. Lett. **91**, 074105 (2007).

37  X. Z. Qin, W. L. Wang, and Z. B. Li, J. Vac. Sci. Technol. B **29**, 031802 (2011).

[38] X. Z. Qin, W. L. Wang, N. S. Xu, Z. B. Li, and R. G. Forbes, Proc. R. Soc. A **467**, 1029 (2011).

[39] T. Mori, M. Hiramatsu, K. Yamakawa, K. Takeda, and M. Hori, Diam. Relat. Mater. **17**, 1513 (2008).

[40] Y. H. Wu, P. W. Qiao, T. C. Chong, and Z. X. Shen, Adv. Mater. **14**, 64 (2002).

[41] M. Hiramatsu and M. Hori, *Carbon Nanowalls* (Springer-Wien, New York, 2010), p. 119.

[42] Z. M. Xiao, J. C. She, S. Z. Deng, Z. K. Tang, Z. B. Li, J. M. Lu, and N. S. Xu, Acs Nano **4**, 6332 (2010).

[43] S. Vadukumpully, J. Paul, and S. Valiyaveettil, Carbon **47**, 3288 (2009).

[44] A. T. T. Koh, Y. M. Foong, L. K. Pan, Z. Sun, and D. H. C. Chua, Appl. Phys. Lett. **101**, 183107 (2012).

[45] W. Takeuchi, H. Kondo, T. Obayashi, M. Hiramatsu, and M. Hori, Appl. Phys. Lett. **98**, 123107 (2011).

[46] C. D. Mee and E. E. Daniel, *Magnetic Recording* (McGraw-Hill, Inc., New York, 1987), Vol. 1: Technology.

[47] Y. H. Wu, Y. Wang, J. Y. Wang, M. Zhou, A. H. Zhang, C. Zhang, Y. J. Yang, Y. N. Hua, and B. X. Xu, Aip Advances **2**, 012132 (2012).

FIG. 1. SEM image for FE measurements on CNW/Cu (a) and etched single-layer graphene on Cu (c) and schematic of the CNW (b) and graphene sample (d). Insets of (a) and (c) are SEM images of the CNW/Cu and etched single-layer graphene sample after all the FE measurements, respectively.

FIG.2. Typical I-E and F-N plots: (a) $I - E$ plots for SLG/Cu, (b) and (c) F-N plots for CVD SLG/Cu, (d) F-N plots for CNW/Cu, and (e) F-N plots for CNW/SiO$_2$. Figures beside the curves are cathode-anode distance (d) in nm.

FIG. 3. (a) Experimental (symbols) and calculated (solid line) enhancement factor as a function of cathode-anode distance for three different types of 2D carbon samples. Inset shows the simulated z-component of the total electric field strength normalized by the global field around the gap region. The rectangular block at the center is the 2D carbon emitter; (b) Experimental data of this study (data in dotted circle) plotted together with the data reported in literature for both localized (unfilled triangle) and large-area FE studies (unfilled diamond) on different kinds of 2D carbon. The dotted line is the average of the reported data from large-area studies; (c) Calculated dependence of the enhancement factor of 2D emitter on normalized sample-anode distance (d/t) at x = 0. Inset shows the schematic of the model.

**(a)** 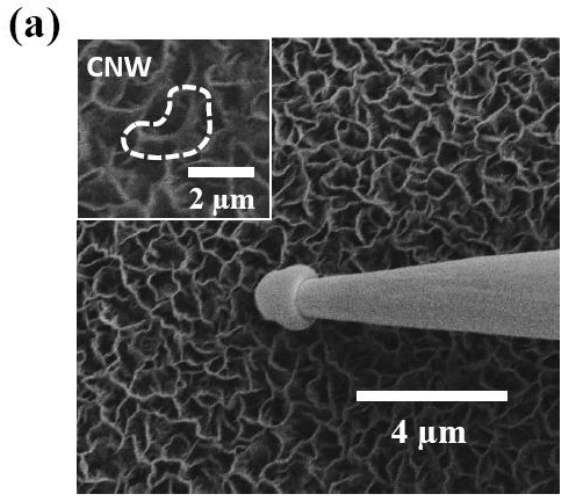

**(b)** 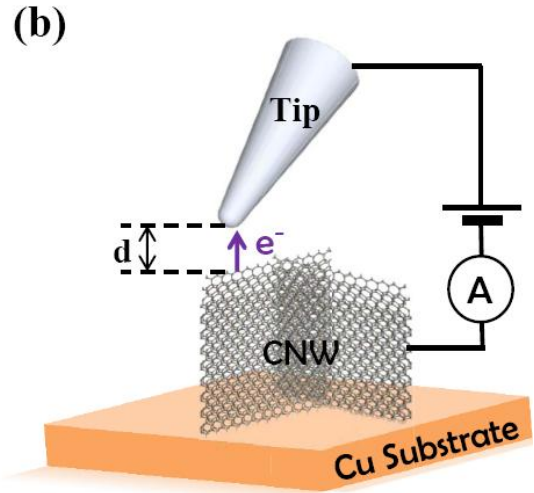

**(c)** 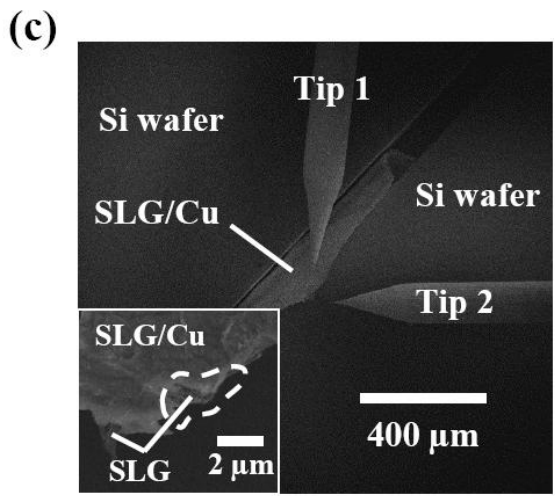

**(d)** 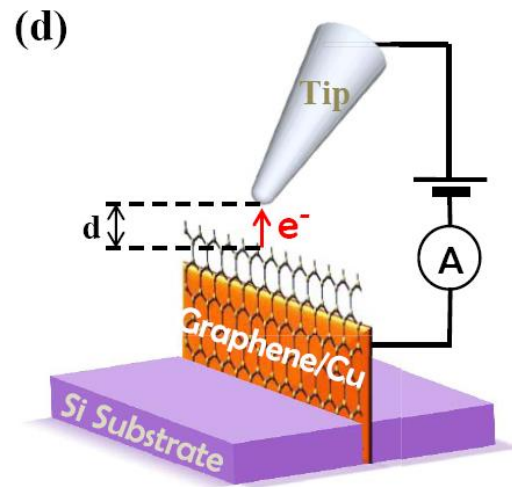

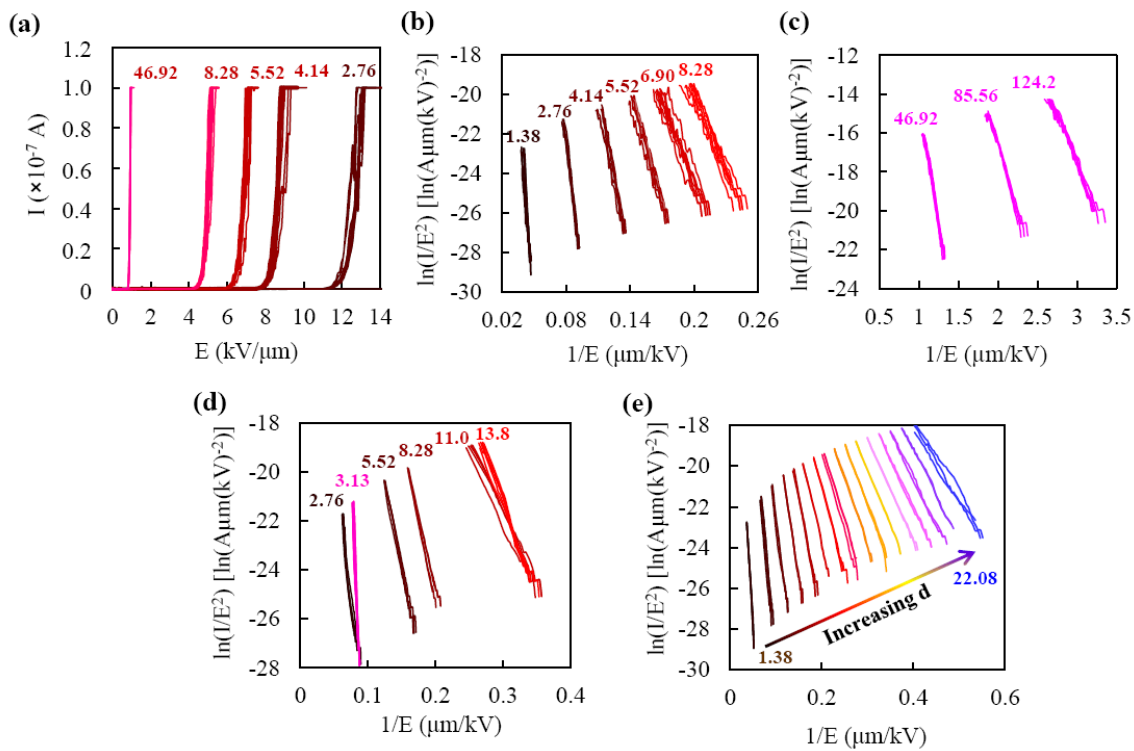

**(a)**

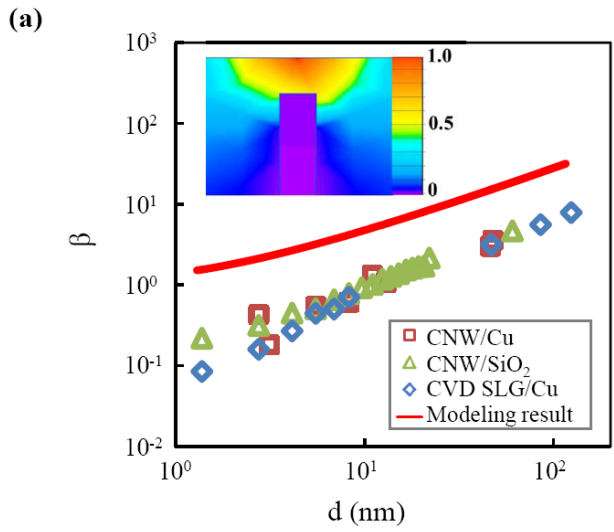

**(b)**

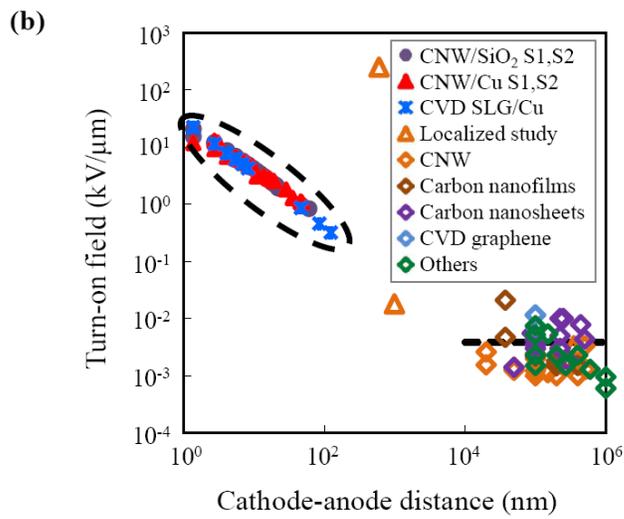

**(c)**

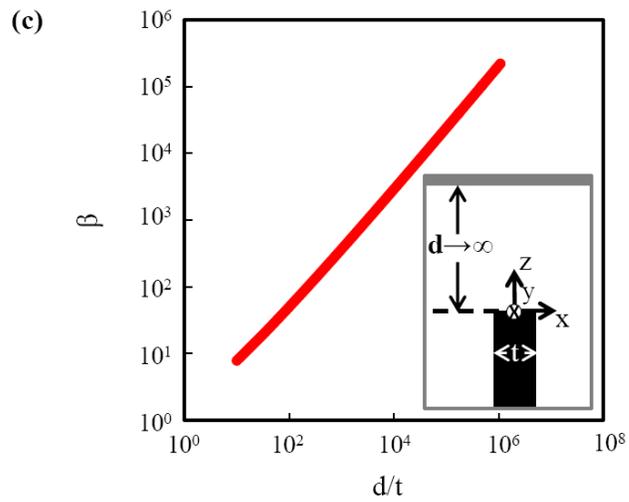